# Evidence of topological edge states in a superconducting nonsymmorphic nodal-line semimetal

L. X. Xu[1,2,3]*, Y. Y. Y. Xia[1,4]*, S. Liu[1,3,5]*, Y. W. Li[1], L.Y.Wei[1], H. Y. Wang[1], C. W. Wang[1,2,3], H. F. Yang[1], A. J. Liang[1], K. Huang[1], T. Deng[1,2,3], W. Xia[1,4], X. Zhang[1], H. J. Zheng[1], Y. J. Chen[6], L. X. Yang[6], M. X. Wang[1], Y. F. Guo[1,4‡], G. Li[1,4§], Z. K. Liu[1,4†], Y. L. Chen[1,4,6,7]

[1]*School of Physical Science and Technology, ShanghaiTech University, CAS-Shanghai Science Research Center, Shanghai 200031, China*
[2] *State Key Laboratory of Functional Material for Informatics, Shanghai Institute of Microsystem and Information Technology, Chinese Academy of Sciences, Shanghai 200050, China*
[3]*University of Chinese Academy of Sciences, Beijing 100049, China*
[4]*ShanghaiTech Laboratory for Topological Physics, Shanghai 201210, China*
[5]*Shanghai Institute of Optics and Fine Mechanics, Chinese Academy of Sciences, Shanghai 201800, China*
[6]*State Key Laboratory of Low Dimensional Quantum Physics, Department of Physics, Tsinghua University, Beijing 100084, China*
[7]*Department of Physics, University of Oxford, Oxford, United Kingdom*

*These authors contributed equally to this work

‡ guoyf@shanghaitech.edu.cn

§ligang@shanghaitech.edu.cn

†liuzhk@shanghaitech.edu.cn

**Topological materials host fascinating low dimensional gapless states at the boundary. As a prominent example, helical topological edge states (TESs) of two-dimensional topological insulators (2DTIs) and their stacked three-dimensional (3D) equivalent, weak topological insulators (WTIs), have sparked research enthusiasm due to their potential application in the next generation of electronics/spintronics with low dissipation. Here, we propose layered superconducting material CaSn as a WTI with nontrivial $Z_2$ as well as nodal line semimetal protected by crystalline non-symmorphic symmetry. Our systematic angle-resolved photoemission spectroscopy (ARPES) investigation on the electronic structure exhibits excellent agreement with the calculation. Furthermore, scanning tunnelling microscopy/spectroscopy (STM/STS) at the surface step edge shows signatures of the expected TES. These integrated evidences from ARPES, STM/STS measurement and corresponding *ab initio* calculation strongly support the existence of TES in the non-symmorphic nodal line semimetal CaSn, which may become a versatile material platform to realize multiple exotic electronic states as well as topological superconductivity.**

## I. INTRODUCTION

One of the most interesting consequences of nontrivial topology in condensed matter systems is the bulk-boundary correspondence, which requires topology protected gapless low dimensional states at the boundary of the parent topological material [1]. Renowned examples include the topological surface states of three-dimensional (3D) strong topological insulators (STIs) [2] and Majorana edge states of topological superconductors (TSCs) [3]. Among them, helical topological edge states (TESs) of two-dimensional (2D) topological insulators (2DTIs) and their stacked 3D equivalent, weak topological insulators (WTIs) have attracted intensive research interest due to their importance to fundamental physics as well as potential application in the next generation of electronics/spintronics with low energy consumption [4-12]. However, TESs are rarely experimentally confirmed, either on the boundary of a 2DTI that usually requires sophisticated device fabrication (e.g. HgTe quantum wells [4], InAs/GaSb quantum wells [5], epitaxially grown Bismuthene on SiC [6] and $WTe_2$ grown on graphene [7,8]) or on the step edge of a WTI (e.g. $Bi_{14}Rh_3I_9$ [9], $ZrTe_5$ [10,11] and $HfTe_5$ [12]).

WTIs are a distinct topological class from STIs in the 3D topological insulators (3DTIs) family[13,14], as shown in Fig. 1a. STIs, such as $Bi_2Se_3$ [15] and $Bi_2Te_3$ [2] exhibit an odd number of surface Dirac cones on all surfaces, whereas WTIs host an even number of inverted band gaps and contain less dispersive TSSs only on the side surfaces [16]. It could also be viewed as stacks of weakly coupled 2DTIs or quantum spin Hall insulators where the electrons at the edge with opposite spins counter-propagating thus immune to back-scattering when time-reversal symmetry is preserved [14,17,18].

In the present study, we suggest layered material CaSn as a new member of WTI with nontrivial $Z_2 = (0; 110)_{primitive}$. *Ab initio* calculation confirms that the naturally cleaved (010) surface is topologically "dark" while TSSs are hosted on the side surfaces. Remarkably, unlike other WTIs with a small energy gap (e.g. $Bi_{14}Rh_3I_9$ [9], $Bi_4I_4$ [19], $ZrTe_5$ [10,11] and $HfTe_5$ [12]), CaSn exhibits Dirac nodal lines protected by non-symmorphic crystalline symmetry which cross the Fermi level thus resulting in semimetallic nature. Systematic angle-resolved photoemission spectroscopy (ARPES) study shows great consistency with the DFT calculation on the band structure. As expected, an additional scanning tunneling spectroscopy (STS) peak near $E_b$ = 220 meV in close vicinity of the step edge compared to the bulk indicates the existence of TES in CaSn. Taken together with these comprehensive evidences from experimental and theoretical aspects, our discovery suggests CaSn as a flexible material platform to exhibit multiple intriguing electronic states including both bulk Dirac line node and TES on the natural stepping edge. Further, as the transport measurement suggests CaSn superconducts below $T_c$ = 4.1 K, the coexistence of topological electronic structure and superconductivity makes CaSn a new platform to investigate topological superconductivity.

## II. EXPERIMENTS AND METHODS

High-quality CaSn single crystals were grown by the melting method. High-purity pieces of Ca (99.95) and pills of Sn (99.999) were mixed with an atomic ratio of Ca:Sn =1:1, then placed in an aluminum oxide crucible and sealed in a silica tube under vacuum. The tube was heated to 1050 °C, kept for more than 20 hours, and then slowly cooled to 800 °C at a rate of 3 °C/h. After that, the tube was quickly cooled to room temperature. At last, the air-sensitive CaSn single crystals were obtained in the aluminum oxide crucible in an Ar-filled glove for ARPES and STM/STS measurements.

The crystallinity and the lattice parameters of the crystals were evaluated by a single-crystal X-ray diffractometer equipped with a Mo K$\alpha$ radioactive source ($\lambda$ = 0.71073 Å). Transport properties were carried out using Physical Properties Measurement System (PPMS) and Magnetic Property Measurement System (MPMS) of Quantum design. The temperature dependent resistance was measured from 2 to 300 K by a four-probe configuration. The temperature dependent magnetization M(T) was measured in the zero-field cooling (ZFC) and field-cooling (FC) mode in the temperature range of 2.5-300 K. Due to the sensitivity of the crystal to air, a layer of paraffin was coated on the crystal to isolate the air for magnetization measurements.

ARPES measurements were performed at SIS-HRPES beamline at Swiss Light Source Synchrotron and BL09U "Dreamline" of the Shanghai Synchrotron Radiation Facility (SSRF) with a Scienta DA30 analyzer. The measurement sample temperature was 10 K and vacuum better than $1.2 \times 10^{-10}$ Torr, respectively. The angle/overall energy resolution was 0.1°/<15 meV, respectively.

STM characterizations were carried out in ultrahigh vacuum (UHV) at ~77 K. The base pressure was $2.5 \times 10^{-10}$ Torr. The CaSn single crystal was cleaved *in situ* in UHV at room temperature and quickly transferred to the sample stage kept at 77 K for STM/STS measurements. PtIr tips were used for both imaging and tunneling spectroscopy, which were all decorated and calibrated on the surface of silver islands grown on p-type Si (111)-7×7. Lock-in technique was employed to obtain dI/dV curves. A modulation voltage of 5 mV at 997.233 Hz was applied to the sample alongside the normal DC sample bias.

The first-principles calculations were carried out within the framework of density functional theory (DFT) as implemented in the Vienna *ab initio* simulation package (VASP) [20]. The Heyd-Scuseria-Ernzerhof (HSE06) hybrid exchange functional [21] was taken, and a *k*-mesh of 7×3×8 for conventional cell and a *k*-mesh of 6×6×9 for the primitive cell are adopted. The crystal structures are optimized before the self-consistent field procedure. The SOC was considered self-consistently. The *s*, *p*, *d* orbitals of Ca and Sn were used to construct the maximally localized Wannier functions [22] via the VASP2WANNIER90 [23] interfaces, which were then used to calculate the LDOS by applying the iterative Green's function approach [24].

## III. RESULTS

CaSn crystalizes in the same base-centered orthorhombic space group (*Cmcm*, No. 63) as the canonical WTI ZrTe$_5$ [10,11], as illustrated in Fig. 1b. The crystal consists of CaSn layers stacking along the [010] direction (or [1$\bar{1}$0] using the primitive vector convention, see Fig. S3) with in-plane lattice constants *a* = 4.81 Å, *c* = 4.35 Å and out-of-plane lattice constant *b* = 11.54 Å. Each layer contains two sublayers of CaSn monolayer consisting of rectangle networks of Ca and Sn nested with each other. The crystal space group contains non-symmorphic symmetry operations including both screw rotation and glide mirror, as elaborated in Supplementary Information, part 1[25]. High-quality CaSn single crystals are synthesized (see Method section) and characterized by X-ray diffraction and core-level spectroscopy (Fig. 1c and Supplementary Information, part 2[25]).

Since the crystal is centrosymmetric, the $Z_2$ topological invariant can be calculated via the product of parities at all time-reversal invariant momenta [26], which confirms the WTI phase ($Z_2$ = (0; 110)$_{primitive}$, more details can be found in Supplementary Information, part 3[25]). Further calculation shows that the top (010) surface is topologically "dark" with Fermi surface (FS) consisting of bulk states whereas the side (001) surface is topological nontrivial hosting TSSs (Fig. 1d).

Figure 1e presents the temperature dependence of resistance and magnetization under zero-field and field cooling. The superconducting transition at $T_C$ = 4.1 K is commonly reported in II$_A$-Sn compounds (CaSn$_3$ [27,28], SrSn$_3$ [29],SrSn$_4$ [30],BaSn$_3$ [31] and BaSn$_5$ [32]). The magnetization as a function of magnetic fields shows a clear hysteresis below $T_C$ (Fig. 1f), which indicates a typical type-II superconducting behavior in contrast to the type-I superconductivity of element Sn [33]. However, we cannot rule out the possibility of extrinsic superconductivity caused by Sn on the degraded surface [34].

Systematic ARPES measurement is carried out to map out the general bulk band structure in the whole 3D Brillouin zone (BZ), as presented in Fig. 2a. Using the tunable photon energy of the synchrotron radiation, we measured the band dispersion along $k_y$ (normal direction of (010) surface), which exhibits clear periodicity (broader range photon-energy dependent measurement can be found in Supplementary Information, part 4[25]) however weakly disperses, indicating relatively weak interlayer coupling and quasi-2D nature.

Figure 2b presents the comparison between calculated and ARPES measured band dispersion along high symmetry momentum directions Z-Γ-X-A-Z (indicated by the green lines at $k_y$ = 0 plane in Fig. 2a), Z-T ($k_y$ direction along the BZ boundary) and T-Y-X$_1$-A$_1$-T (indicated by the red lines at $k_y$ = π plane in Fig. 2a), showing excellent agreement. Non-symmorphic crystalline symmetry enforce the band degeneracy on the BZ boundary ($k_z$ = π, A-Z-T-A$_1$ plane) under negligible spin-orbit coupling (SOC). When SOC is taken into account, the band degeneracy is lifted, however, the Dirac nodal line along Z-T remains intact, which is clearly observed in the Dirac band dispersion evolution shown in Fig. 2c and Fig. S5. The detailed analysis of band degeneracy protected by non-symmorphic symmetry with and without SOC is elaborated in Supplementary Information, part 6[25].

ARPES measurement is performed on the naturally cleaved (010) surface, as illustrated in Fig. 3a. The FS consists of an elliptical hole-like pocket around $\bar{\Gamma}$ and an electron-like pocket near $\bar{Z}$ (see Fig. 3b and 3d). The side-by-side comparison between ARPES measured and calculated band dispersion and energy contours are presented in Fig. 3c and 3d, respectively, exhibiting perfect consistency and indicating all electronic states of bulk origin.

To unravel the expected TES in the WTI phase of CaSn, we acquire STS near and away from the step edge on the (010) surface, which is proportional to the local density of states (LDOS), as shown in Fig. 4a. The surface spectrum far away from the step (bulk states) clearly shows the metallicity of the sample with a finite density of states in the whole energy range detected, consistent with ARPES results (details of the comparison can be found in Supplementary Information, part 7[25]). In contrast to the featureless bulk states in the energy range (-0.4 eV, 0 eV), the STS on the step edge exhibits an additional peak feature near $E = -0.22$ eV and higher DOS in the negative bias range, which is attributed to the TES. Figure 4b maps the spatial distribution of LDOS at different sample bias near a straight step edge. The DOS near the terrace is significantly enhanced at -0.5 eV and -0.2 eV, while slightly suppressed under positive bias (0.1 eV and 0.2 eV), which is a clear evidence for the existence of TES, resembling the behavior of the step edge of WTe$_2$ [35] and ZrTe$_5$ [11]. STS line scan perpendicular to a step edge (Fig. 4c) indicates that the observed TES is closely localized along the edge, with a characteristic length $\xi = 2$ nm (the fitting method is presented in Supplementary Information, part 8[25]). Such decay length falls in the same range of several nanometres as the TES observed in other WTIs and 2DTIs [9-12].

The existence of the TES on the stepping edge of the (010) surface is consistent with the picture that WTI can be regarded as weakly coupled 2D QSH layers stacked along specific directions [35]. In the present case, CaSn single crystals are weakly coupled layers stacked along the [010] direction, or $b_1$-$b_2$ using the base vectors of the primitive cell (see Fig. S3), consistent with our calculated 3D $Z_2$ number $(v_0; v_1 v_2 v_3) = (0;110)_{\text{primitive}}$.

## IV. DISCUSSION AND SUMMARY

CaSn is a reminiscence of other interesting topological materials. It shares many similarities with WTI family MSiX (M = Zr, Hf; X = O, S, Se, Te), which also hosts non-symmorphic symmetry protected Dirac nodal lines [36] and TES in its 2D limit [37]. One of the intriguing properties of these compounds is that they have a similar crystal structure as the well-known iron-pnictide superconductor LiFeAs [37], which promises a possible combination of superconductor and 2DTI. Another relative compound is $WTe_2$, which exhibits TES on the step edge [35] and quantum spin hall effect in its monolayer form [7,38]. The superconductivity can be successfully induced in monolayer $WTe_2$ by epitaxially grown on the superconducting $NbSe_2$ [39]. Compared to those compounds, WTI phase, non-symmorphic symmetry protected Dirac nodal line, and intriguing conducting edge states coexist in CaSn, making it a unique platform to realize multiple exotic electronic states. Furthermore, we notice that the in-plane constants of CaSn roughly match with that of semiconducting substrate InSb [40] and type-II BCS superconducting substrate Nb [41] (details can be found in Supplementary Information, part 9[25]), which imply promising epitaxial grown functional heterostructure applications.

Although we provide compelling evidence of the exotic TES on the natural step edge of the layered crystal CaSn, many questions remain open and some areas of the topic are still unexplored which require further investigation. For example, future research can study the response of the TES under a magnetic field [11] and investigate the TES evolution when CaSn crystal is thinned down to the 2D limit. The nature of observed superconductivity, as well as its interplay between the TES would establish a playground for 1D topological superconductivity, which can serve as a key ingredient for the realization of topological quantum computation.

To summarize, we have presented a comprehensive understanding of the 3D electronic structure of CaSn. Systematic ARPES investigation clearly confirms the non-symmorphic symmetry protected Dirac line node along the Z-T direction. Direct observation of the conducting edge state by STM/STS measurements, the band structure measured by ARPES and corresponding *ab initio* calculations are consistent and provide strong evidences of the WTI phase and accompanying TES, which suggest CaSn as a flexible material platform to realize rich exotic electronic states as well as topological superconductivity.

**ACKNOWLEDGMENTS**

We thank Dr. Yangyang Lv for the assistance and discussion. The work was supported by the National Key R&D program of China (Grants No. 2017YFA0305400 and No.2017YFE0131300) and the National Natural Science Foundation of China (Grant No. 92065201). G.L acknowledges the support from the Strategic Priority Research Program of Chinese Academy of Sciences (Grant No. XDA18010000) and Shanghai Technology Innovation Action Plan 2020-Integrated Circuit Technology Support Program (Project No. 20DZ110060X). Y.F.G acknowledges the support from the Program for Professor of Special Appointment (Shanghai Eastern Scholar). Y.W.L acknowledges the support from the International Postdoctoral Exchange Fellowship Program (Talent-Introduction Program, Grant No. YJ20200126). H.F.Y acknowledges the support from Shanghai Sailing Program (Grant No. 20YF1430500) and National Natural Science Foundation of China (Grant No. 12004248). L.X.X acknowledges the support from the UCAS Joint PhD Training Program.

Figure 1

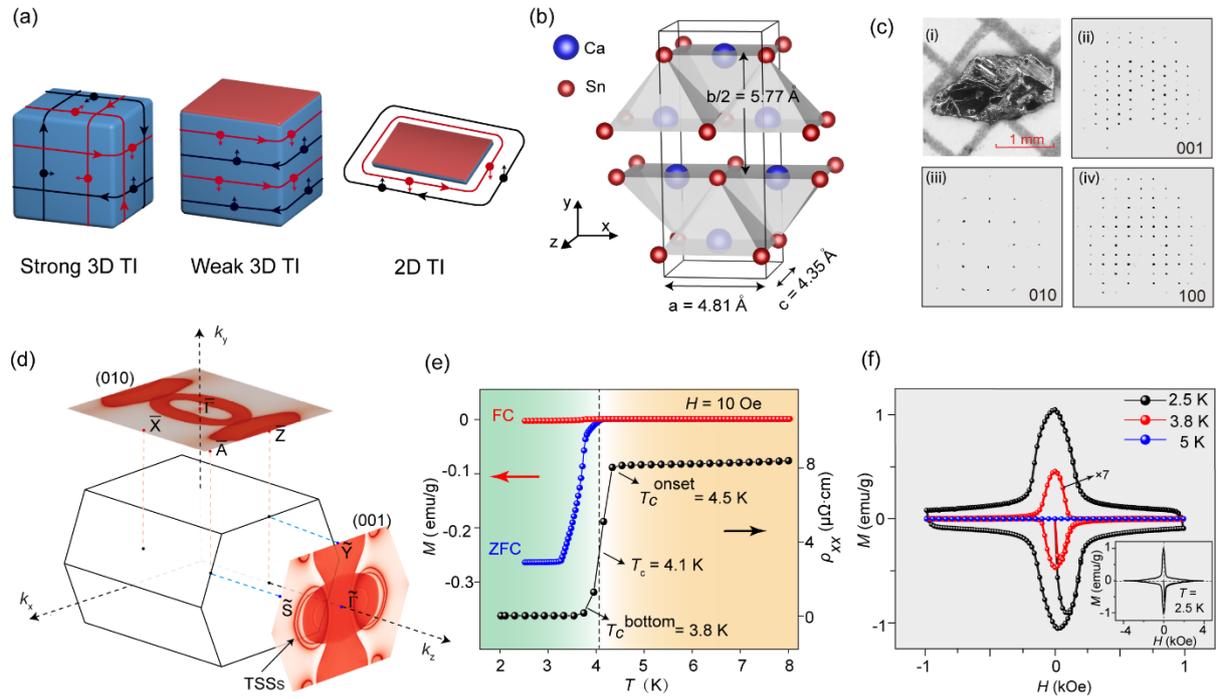

FIG. 1. General characterizations of CaSn. (a) Real-space illustration of the surface state in the strong 3D topological insulator (i), weak 3D topological insulator (ii) and 2D topological insulator (iii). Spins-up (black arrows) and spin-down (red arrows) electrons move in opposite directions. (b) Schematic illustration of the crystal structure of CaSn, showing two adjacent CaSn layers. (c) Optical image of the high-quality CaSn single crystal (i). X-ray diffraction patterns from the (001) (ii), (010) (iii), (100) (iv) surface of CaSn. (d) Three-dimensional Brillouin zone with calculations of surface-projected energy contours on the (001) side surface and (010) top surface. (e) Magnetization and resistance as functions of temperature, show a sharp superconducting transition at $T_C$ = 4.1 K. (f) Magnetic moment as a function of magnetic field measured at 2.5 K (black), 3.8 K (red) and 5 K (blue). Inset: Hysteresis loop of magnetization curve at 2.5 K with a large magnetic field range.

Figure 2

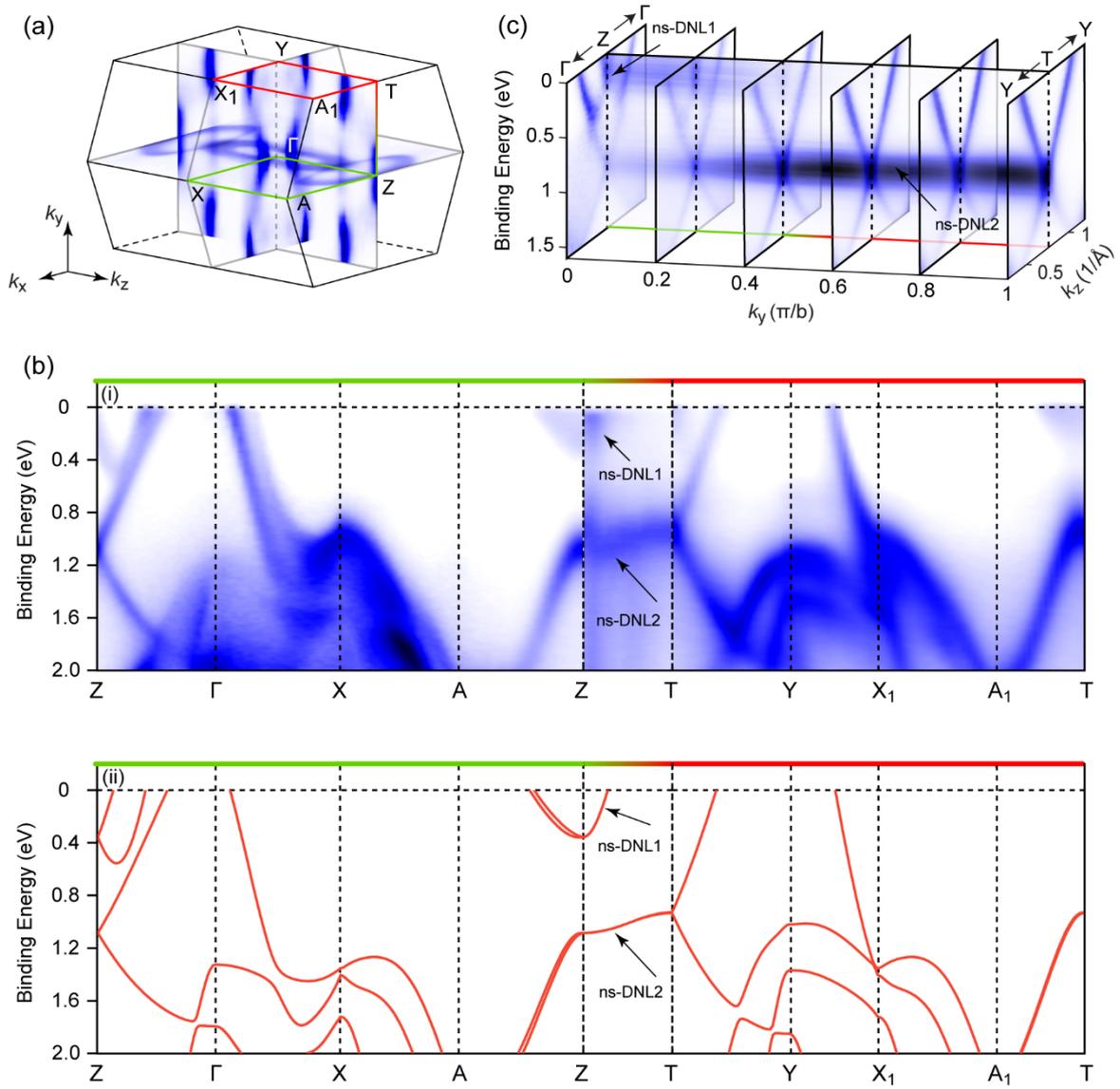

FIG. 2. Overall electronic structures across the 3D Brillouin zone. (a) The 3D BZ shows the Fermi surfaces on the $k_x - k_y$, $k_x - k_z$ and $k_y - k_z$ planes with high symmetry momenta labelled. (b) The measured band dispersions (i) and corresponding calculated bulk structures (ii) along the high-symmetry directions Z-Γ-X-A-Z (indicated by the green lines at $k_y = 0$ plane in a), Z-T ($k_y$ direction along the BZ boundary) and T-Y-$X_1$-$A_1$-T (indicated by the red lines at $k_y = \pi$ plane). (c) Evolution of Dirac band dispersions along $k_y$ (ZT direction). Stack plot of the six cuts with $k_y$ values ranging from 0 to $\pi/b$. Dirac nodal lines are indicated by the black arrows.

Figure 3

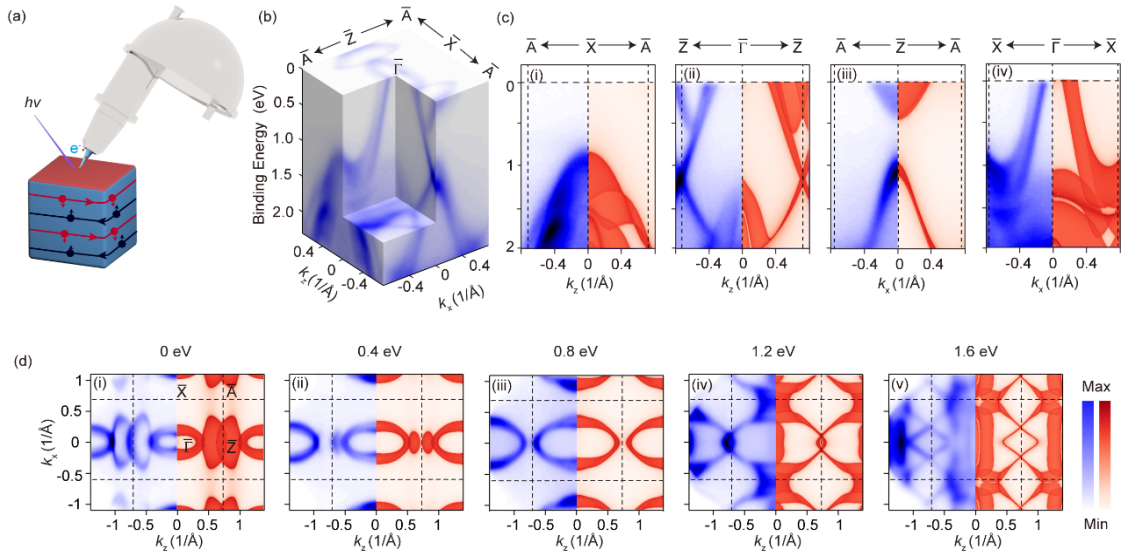

FIG. 3. ARPES measurement on topologically dark (010) surface. (a) The experiment geometry of ARPES measurement on the top surface of a weak 3D topological insulator. (b) The 3D intensity plot of ARPES data on the naturally cleaved (010) surface. (c) Photoemission intensity (left panel) and corresponding *ab initio* calculation (right panel) along the high symmetry $\bar{X}\bar{A}$ (i), $\bar{\Gamma}\bar{Z}$ (ii), $\bar{Z}\bar{A}$ (iii), $\bar{\Gamma}\bar{X}$ (iv) directions. (d) Measured (left panel) and calculated (right panel) constant energy contours (CECs) with five different binding energies $E_b$ = 0 eV (i), 0.4 eV (ii), 0.8 eV (iii), 1.2 eV (iv) and 1.6 eV (v), respectively. The experimental plot has been symmetrized with respect to the $k_x$ =0 plane according to the crystal symmetry. The raw data of ARPES measurement are presented in Fig. S 10[25].

Figure 4

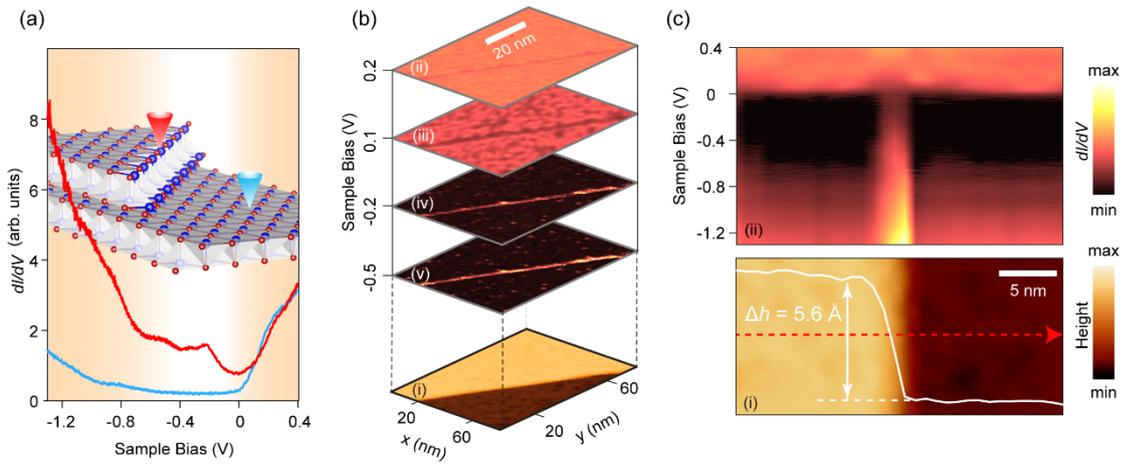

Fig. 4. STM/STS measurements of the edge states. (a) Typical STS measured at the step edge (red curve) and a location away from the edge (blue curve). Inset: Schematics of CaSn (010) termination with a step edge. (b) STM topographic (i), 80 nm × 80 nm, with sample bias $V_s$ = 400 mV, tunneling current $I_s$ = 200 pA. (ii-v) The corresponding $dI/dV$ mapping of the imaged area (i) at different sample bias. (c) (i) STM topography appended with the line profiles along the dashed red arrow, showing a step of 5.6 Å. (ii) Differential conductance along the dotted red line in (i) within a bias range -1.2 V~ +0.4 V.